\documentstyle[12pt,aps,prd,preprint]{revtex}
\begin{document}
\title{Improved Upper Bound to the Entropy of a Charged System. II}
\author{Shahar Hod}
\address{The Racah Institute for Physics, The
Hebrew University, Jerusalem 91904, Israel}
\date{\today}
\maketitle
\begin{abstract}

Recently, we derived an {\it improved} universal upper bound to the
entropy of a {\it charged} system $S \leq \pi (2E b-q^2)/ \hbar$. 
There was, however, some 
uncertainty in the value of the numerical factor which
multiplies the $q^2$ term. In this paper we remove this uncertainty; 
we rederive this upper bound from 
an application of the generalized second law of thermodynamics to a 
gedanken experiment in which an entropy-bearing charged system falls
into a Schwarzschild black hole. A crucial step in the analysis is the
inclusion of the effect of the spacetime {\it curvature} 
on the electrostatic {\it self-interaction} of the charged system.
\end{abstract}
\bigskip

According to the thermodynamical analogy in black-hole physics, the
entropy of a black hole \cite{Beken1,Beken2,Beken3} is given 
by $S_{bh}=A/4 \hbar$, where $A$ is the
black-hole surface area. (We use gravitational units in which $G=c=1$).
Moreover, a system
consisting of ordinary matter interacting with a black hole is widely
believed to obey the {\it generalized second law of thermodynamics} (GSL): 
``{\it The sum of the black-hole entropy and the common (ordinary) 
entropy in the black-hole exterior never decreases}''. This general conjecture
is one of the corner stones of black-hole physics.

It is well known, however, that the validity of the GSL depends on the
(plausible) existence of a universal upper bound to the entropy of a
bounded system \cite{Beken4}: Consider a box filled with matter of proper energy $E$ and 
entropy $S$ which is dropped into a black hole. The energy delivered
to the black hole can be arbitrarily {\it red-shifted} by letting the
assimilation point approach the black-hole horizon. 
If the box is deposited with no radial momentum
 a proper distance $R$ above the horizon, and then allowed to fall in such that

\begin{equation}\label{Eq1}
R < \hbar S/2 \pi E\  ,
\end{equation}
then the black-hole area increase (or equivalently, the increase in 
black-hole entropy) is not large enough to compensate for the decrease
of $S$ in common (ordinary) entropy. Arguing from the GSL, Bekenstein
\cite{Beken4} has proposed the existence of a universal upper bound to the
entropy $S$ of any system of total energy $E$ and effective proper radius $R$:

\begin{equation}\label{Eq2}
S \leq 2\pi RE/\hbar\  ,
\end{equation}
where $R$ is defined in terms of the
area $A$ of the spherical surface which circumscribe the 
system $R=(A/4\pi)^{1/2}$ \cite{Beken4}.
This restriction is necessary for enforcement of the GSL; the
box's entropy disappears but an increase in black-hole entropy occurs
which ensures that the GSL is respected provided $S$ is bounded 
as in Eq. (\ref{Eq2}). Evidently, this universal upper bound is 
a {\it quantum} phenomena (the upper bound goes to infinity as $\hbar \to
0$). This provides a striking illustration of the fact that the GSL is
intrinsically a quantum law. The universal upper bound
Eq. (\ref{Eq2}) has the status of a supplement to the second law; the
latter only states that the entropy of a closed system tends to a
maximum without saying how large that should be.

Other derivations of the universal upper 
bound Eq. (\ref{Eq2}) which are based on black-hole physics have been
given in \cite{Zas1,Zas2,Zas3,LiLiu}. Few pieces of evidence exist concerning the
validity of the bound for self-gravitating systems
\cite{Zas1,Zas2,Sorkin,Zurek}. However, the universal bound Eq. (\ref{Eq2})
is known to be true independently of black-hole physics for a variety
of systems in which gravity is 
negligible \cite{Beken5,Beken6,BekenSchi,SchiBeken,BekenGuen}.

We noted \cite{Hod1,Hod2}, however, that there is one disturbing feature of the universal 
bound Eq. (\ref{Eq2}): Black holes conform to the bound \cite{Beken4};
however, it is only the Schwarzschild black hole which actually
{\it saturates} the bound. This uniqueness of
the Schwarzschild black hole (in the sense that it is the {\it only} 
black hole which have the maximum entropy allowed by quantum theory and
general relativity) is somewhat 
disturbing. Recently, Hod \cite{Hod1} derived an
(improved) upper bound to the entropy of a {\it spinning} system and proved 
that {\it all} electrically neutral Kerr black holes 
have the {\it maximum} entropy allowed by quantum theory 
and general relativity. The unity of physics (and of black holes in
particular) motivates us to look for an improved upper bound to
the entropy of a {\it charged} system.

Moreover, the plausible existence of an upper bound stronger than
Eq. (\ref{Eq2}) on the entropy of a charged system 
has nothing to do with black-hole physics; a part of the energy of the
electromagnetic field residing outside the charged system seems to be irrelevant 
for the system's statistical properties. This reduce the phase space
available to the components of a charged system. Evidently, an improved
upper bound to the entropy of a charged system must {\it decrease}
with the (absolute) value of the system's charge. 
However, our simple argument cannot yield the exact dependence of the
entropy bound on the system's parameters: its energy, charge, and proper radius. 

It is black-hole physics (more precisely, the GSL) which yields a
concrete expression for the universal upper bound; 
recently, we have derived an {\it improved} universal upper bound to the
entropy of a {\it charged} system $S \leq \pi (2E b-q^2)/ \hbar$ \cite{Hod2}. 
There was, however, some
uncertainty in the value of the numerical factor which
multiply the $q^2$ term. In this paper we remove this uncertainty. 

We consider a charged body of rest mass $\mu$, charge
$q$, which is dropped into a Schwarzschild black hole.
The equation of motion of a charged body on a Schwarzschild 
background is a quadratic equation
for the conserved energy $E$ (energy-at-infinity) of the body \cite{Carter}

\begin{equation}\label{Eq3}
r^4 E^2 -
\Delta(\mu^2 r^2 +{p_{\phi}}^2)- (\Delta p_r)^2=0\  ,
\end{equation}
where $\Delta=r^2-2Mr$.
The quantities $p_{\phi}$ and $p_r$ are the conserved 
angular momentum of the body and its
covariant radial momentum, respectively. 

The conserved energy $E$ of a body having a radial turning 
point at $r=r_{+}+ \xi$ \cite{note1} (for $\xi \ll r_{+}$ 
where $r_+=2M$ is the location
of the black-hole horizon) is given by Eq. (\ref{Eq3})

\begin{equation}\label{Eq4}
E=\sqrt{\mu^2+p_{\phi}^2/{r_+}^2} {(\xi/r_+)}^{1/2}[1+O(\xi /r_{+})]\  .
\end{equation}
This expression is actually the effective potential (gravitational 
plus centrifugal) for given 
values of $\mu$ and $p_{\phi}$.
It is clear that it can be minimized by taking $p_{\phi}=0$ (which
also minimize the increase in the black-hole surface area).

However, the well-known analysis of \cite{Carter} is not complete because
it does {\it not} take into account the effect of the 
spacetime {\it curvature} on the particle's 
electrostatic {\it self-interaction}. The black-hole gravitational
field modifies the electrostatic
self-interaction of a charged particle in such a way that the particle
experience a repulsive (i.e., directed away from the
black hole) self-force. A variety of
techniques have been used to demonstrate this effect
\cite{DeDe,Ber,Mac,Vil,SmWi}. The physical origin of this force 
is the distortion of the charge's long-range Coulomb field by
the spacetime curvature. The contribution of this effect to the
particle's energy is $Mq^2/2r^2$ \cite{SmWi}. 

In order to find the change in black-hole surface area caused by
an assimilation of the body, one should evaluate $E$ 
at the point of capture, a proper distance $b$
outside the horizon. The relevant dimension of the body in our
gedanken experiment is its shortest length. In other words, the entropy
bound is set by the smallest body's dimension (provided $b \gg
\hbar / E$ \cite{Beken6}). This conclusion is supported by numerical
computations \cite{Beken5} for neutral systems. Thus, we should
evaluate $E$ at $r=r_{+}+ \delta (b)$, where $\delta(b)$ is 
determined by 

\begin{equation}\label{Eq5}
\int_{r_{+}}^{r_{+}+ \delta (b)} (1-2M/r)^{-1/2} dr = b\  .
\end{equation}
Integrating Eq. (\ref{Eq5}) one obtains (for $b \ll r_{+}$)

\begin{equation}\label{Eq6}
\delta (b)=b^2/8M\  ,
\end{equation}
which implies (to leading order in b/M)

\begin{equation}\label{Eq7}
E=(2\mu b+q^2)/8M\  .
\end{equation}

An assimilation of the charged body results in a change $\Delta M=E$ in the
black-hole mass and a change $\Delta Q=q$ in its charge.
The relation $A=4\pi[M+(M^2-Q^2)^{1/2}]^2$ implies that 
(for $Q=0$) $\Delta A=8\pi[4M \Delta M-(\Delta Q)^2]$ (terms of order
$(\Delta M)^2$ are negligible for $b \ll M$). Thus, taking cognizance of
Eq. (\ref{Eq7}) we find 

\begin{equation}\label{Eq8}
(\Delta A)_{min}=4\pi(2\mu b -q^2)\  ,
\end{equation}
which is the {\it minimal} black-hole area increase for given values of the
body's parameters $\mu, q,$ and $b$.
Assuming the validity of the GSL, one can derive an upper bound to 
the entropy $S$ of an arbitrary system of proper energy $E$, charge
$q$ and circumscribing radius $R$ (by definition, $R \geq b$):

\begin{equation}\label{Eq9}
S \leq \pi (2ER-q^2)/ \hbar\  .
\end{equation}
It is evident from the minimal
black-hole area increase Eq. (\ref{Eq8}) that in order for the GSL to
be satisfied $[(\Delta S)_{tot} \equiv (\Delta S)_{bh} -S \geq 0]$,
the entropy $S$ of the charged system must be bounded as in
Eq. (\ref{Eq9}). This upper bound is {\it universal} in the sense that it
depends only on the {\it system's} parameters 
(it is {\it independent} of the black-hole mass which was used to
derive it).

This improved bound is very appealing from a black-hole physics point
of view \cite{Hod2}: consider a {\it charged} Reissner-Nordstr\"om
black hole of charge $Q$. Let its energy be $E$; then its 
surface area is given by $A=4 \pi {r_{+}}^2 =4\pi (2E r_+-Q^2)$. 
Now since $S_{bh}=A/4\hbar$,
$S_{bh}=\pi (2Er_+-Q^2) /\hbar$, which is the {\it maximal} 
entropy allowed by the upper bound Eq. (\ref{Eq9}). Thus, {\it all}
Reissner-Nordstr\"om black holes {\it saturate} the bound. This proves
that the Schwarzschild black hole is {\it not} unique from a black-hole entropy point of
view, removing the disturbing feature of the entropy bound
Eq. (\ref{Eq2}). This is {\it precisely} the kind of universal upper 
bound we were hoping for !

Evidently, systems with negligible self-gravity (the charged system in our
gedanken experiment) and systems with maximal gravitational effects 
(i.e., charged black holes) both satisfy the upper bound
Eq. (\ref{Eq9}). Therefore, this bound appears to be of universal
validity. One piece of evidence exist concerning the
validity of the bound for the specific example of a system composed 
of a charged black hole in thermal equilibrium with radiation \cite{Zas2}.

The intriguing feature of our derivation is that it uses a law
whose very meaning stems from gravitation (the GSL, or equivalently
the area-entropy relation for black holes) to derive a universal 
bound which has nothing to do with 
gravitation [written out fully, the entropy bound 
would involve $\hbar$ and $c$, but not
$G$]. This provides a striking illustration of the unity of physics.

In summary, an application of the generalized second law of
thermodynamics to a gedanken experiment in which 
an entropy-bearing charged system falls 
into a Schwarzschild black hole, enables us to 
derive an {\it improved universal upper bound} to the entropy of 
a {\it charged} system \cite{Hod2}. In doing so, 
we removed the former uncertainty regarding the precise 
value of the numerical coefficient which multiply the $q^2$ term.
A crucial step in the analysis is the
inclusion of the influence of the spacetime {\it curvature} 
on the system's electrostatic {\it self-interaction}.

Note added: I have learned that recently Bekenstein and Mayo
[Phys. Rev. D {\bf 61}, 024022 (2000)] analyzed the same problem, and independently 
obtained the 
universal upper bound (which was already derived in \cite{Hod2}).

\bigskip
\noindent
{\bf ACKNOWLEDGMENTS}
\bigskip

It thank Jacob D. Bekenstein and Avraham E. Mayo for helpful discussions. 
This research was supported by a grant from the Israel Science Foundation.

\end{document}